\documentstyle[epsf,psfig]{l-aa}

\thesaurus{11.03.4 Coma; 11.03.5; 11.12.2; 11.16.1}

\title{A photometric catalogue of the Coma cluster core 
\thanks{Based on observations collected at the 
Canada-France-Hawaii telescope, operated by the 
National Research Council of Canada, the Centre National de la Recherche
Scientifique of France, and the University of Hawaii}}

 \author {
C.~Lobo \inst{1,2}
\and
 A.~Biviano \inst{3,4}
\and
  F.~Durret \inst{1,5}
\and
  D.~Gerbal \inst{1,5}  
\and
 O.~Le F\`evre \inst{5}
\and
 A.~Mazure \inst{6} 
\and
 E.~Slezak \inst{7}
}
\offprints{C.~Lobo, lobo@iap.fr }
\institute{
	Institut d'Astrophysique de Paris, CNRS, Universit\'e Pierre et 
	Marie Curie, 98bis Bd Arago, F-75014 Paris, France 
\and
	Centro de Astrof\'\i sica da Universidade do Porto, Rua do Campo 
Alegre 823, P-4150 Porto, Portugal
\and
	Istituto T.E.S.R.E., Area della Ricerca del CNR, via Gobetti 101,
	I-40129, Bologna, Italy
\and
	ESA Villafranca Satellite Tracking Station, Apto. 50727, CAM IDT,
	E-28080 Madrid, Spain
\and 
	DAEC, Observatoire de Paris, Universit\'e Paris VII, CNRS (UA 173), 
	F-92195 Meudon Cedex, France 
\and
    	IGRAP, LAS, Traverse du Siphon, Les Trois Lucs, B.P. 8, F-13376 
	Marseille Cedex, France 
\and
    	Observatoire de la C\^ote d'Azur, B.P. 229, F-06304 Nice Cedex 4, 
	France 
}
\date{Received 1996 May 23; accepted ...}
\begin{document}

\maketitle

\begin{abstract}~\footnote{The catalogue is only available in electronic form 
at the CDS via anonymous ftp to cdsarc.u-strasbg.fr (130.79.128.5) or via 
http://cdsweb.u-strasbg.fr/Abstract.html}
We have obtained a mosaic of CCD images of the Coma cluster in the V-band 
covering a region of approximately 0.4 degrees$^2$ around both central 
cluster galaxies NGC~4889 and NGC~4874. An additional frame of $\sim$~90 
arcmin$^2$ was taken of the south-west region around NGC~4839. We derived a 
catalogue of 7023 galaxies and 4096 stars containing positions, central 
surface brightnesses and isophotal V$_{26.5}$ magnitudes. We estimate that 
data is complete up to V~$_{26.5}\sim$~22.5 and the surface brightness 
limiting detection value is $\mu$~$\sim$~24~mag/arcsec$^2$.\\
In this paper we present the catalogue (available in electronic form alone), 
along with a detailed description of the steps concerning the data reduction 
and quality of the computed parameters.

\end{abstract}

\section{Observations} 
\begin{figure}
\centerline{\psfig{file=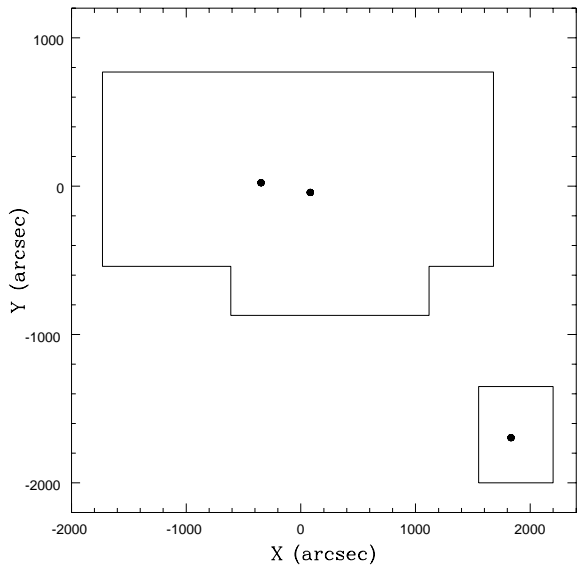,height=8.0cm}}
\caption{Map of the observed areas. Dots indicate the 
positions of the giant 
galaxies NGC~4874 and NGC~4889 in the centre, and NGC~4839 in the south-west 
frame. Coordinates are given relatively to the GMP centre - see 
section~\ref{xy} - located at $\alpha=12^h\ 57.3^m$, $\delta =28^\circ\ 
14.4$' (1950.0). North is up, east is to the left.}
\protect\label{map}
\end{figure}

We have observed at the 3.6m Canada-France-Hawaii Telescope during four 
nights in May 1993 with the MOS-SIS spectrograph (Le~F\`evre et al. 
1994) in the imaging mode. 
The Loral3 CCD, which 
has a 2048~$\times$~2048 pixel format, provides images of 9.7~$\times$~9.4~
arcmin$^2$ (after discarding the vignetting area) -- at the distance of the 
Coma cluster, 10~arcmin correspond to 0.4~h$_{50}^{-1}$~Mpc -- and the pixel 
size is 0.3145~arcsec.
A ``mosaic'' of 21 overlapping images in the V-band was thus obtained 
covering a total field of 
about 0.4 degrees$^2$ centered on the two brightest central galaxies
of Coma (NGC~4874 and NGC~4889). An additional frame was taken of the 
south-west NGC~4839 group. In Fig.~\ref{map} we display the observed regions. 
The exposure time for each image was 3 minutes.
Flat--field frames of the twilight sky were also obtained with 1~second 
exposure time each, as well as a standard star calibration field in M92 
with a 90~second exposure. During the whole run the seeing (as 
estimated by the point spread function of stars in the images) varied from 
0.9 to 1.4 arcsec.

\section{Flat--field, bias subtraction and correction of MOS distortions}
All the data reduction was performed with the IRAF package. Bias and 
flat--field corrections were made in the usual way. We used the twilight 
flat--field rather than a median flat produced from the images because the 
projected density and size of some of the bigger Coma galaxies did not allow 
to obtain a flat--field totally free of residuals.\\
We applied the correction for distortion caused by the MOS camera optics 
(Le F\` evre et al. 1994) that mainly affects the corners of the CCD. 
This is done by running, for each image, the task GEOTRAN that corrects 
the distribution of the photon flux in the image pixels by means of a 
distortion map especially designed for this instrument.\\ 
Hot pixels, cosmics and CCD defects (bad columns, dead pixels,...) 
were flagged by eye inspection of each image, thus completing the 
pre-reduction stage.

\section{Detection of objects}
Objects were automatically detected using the task DAOPHOT/DAOFIND. This task 
performs a convolution with a gaussian having characteristics previously 
chosen taking into account the seeing in each frame (FWHM of the star-like 
profiles in the image) as well as the CCD readout noise and 
gain. Then, objects are identified as the peaks of the convolved image that 
are higher than a given threshold 
above the local sky background (chosen as approximately equal to 5~$\sigma$ 
of the image mean sky flux).
A list of detected objects is thus produced and interactively corrected on 
the displayed image so as to discard spurious objects, add undetected ones 
and dispose of false detections caused by the events flagged in the previous 
section (all of which concerning only a few percent of objects). Notice that 
all objects that were ``hand-added'' to the final list are both very faint 
and very low surface brightness ones, though still visible by eye inspection. 
The completeness of the catalogue is by no means dependent on this correction.
 
\section{Photometry}

\subsection{Choice of the isophotal level and magnitude calibration}\label{cal}
The isophote we selected to measure the flux of the objects, taking into 
account the S/N of the images, corresponds to 26.5 mag$/arcsec^2$. Subsequent 
tests confirmed that this value provides magnitudes very close to total ones~: 
for objects with isophotal magnitude (simply noted as V hereafter) up to 
V~$\sim$~21.0, the difference 
between our measured value and a total magnitude (as estimated by a Kron 
magnitude) is lower than 0.05 magnitudes and the shift is non-systematic. 
In what concerns fainter objects, the isophotal radius seems to be 
overestimating the object radius and we thus measure isophotal magnitudes 
that can be, at most, 0.2 magnitudes brighter than the total estimates.\\
Several standard stars in the M92 star cluster field were used to 
calibrate the photometry. The calibrated magnitudes for these objects, ranging 
from V~$\sim$~14 up to V~$\sim$~19.4, were catalogued by 
Christian et al. (1985).  We measured their fluxes in the image and 
computed the corresponding apparent magnitudes, which were then compared 
to the calibrated ones. By taking into account the different exposure times 
(90~seconds for the M92 image vs. 3~minutes for the Coma frames) we thus 
produced a zero-point calibration constant. All nights were photometric 
and, in such conditions, the zero point variation throughout one night and 
from night to night is less than 0.1 magnitude (see section~\ref{zeropoint}). 
So, we have assumed this same calibration zero-point to be valid for all Coma 
frames. The airmass term is negligible for all frames.\\
This relatively large photometric uncertainty is probably mostly due to the 
lack of a color term in the photometrical calibration: the absence of a 
second filter makes it impossible to compute such a correction term. 

\subsection{Data reduction}

We used the package developed by Olivier Le F\`evre (Le F\`evre et al. 
1986, Lilly et al. 1995) to reduce 
the data and obtain a catalogue with (x,y) position, isophotal radius and 
magnitude within the 26.5 isophote, and central surface brightness for more 
than 11000 stars and galaxies. 
This software has the advantage of having been created especially for this 
kind of photometry and extensively tested on MOS CFHT observations.
\\

\subsection{Star-galaxy separation}
\begin{figure}
\centerline{\psfig{file=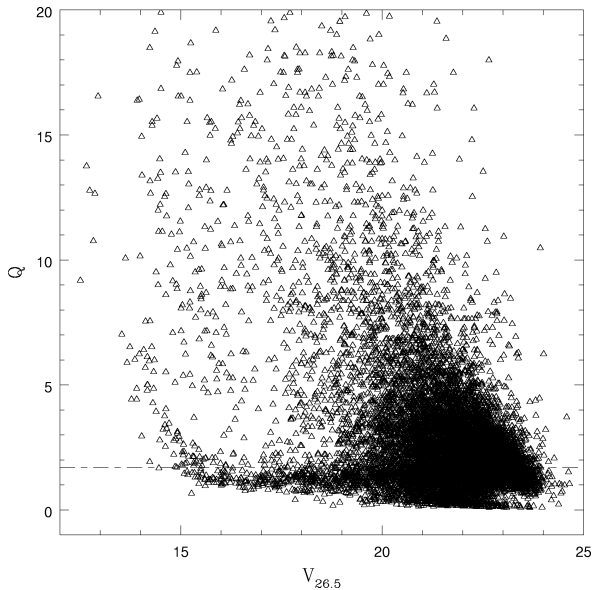,height=8.0cm}}
\caption{Normalized compactness parameter Q vs. isophotal magnitude for all 
observed objects. The dashed line indicates the Q$_{sep}$~=~1.7 value 
determined 
to separate stars (below the line) from galaxies (above the line).}
\protect\label{sep}
\end{figure}

\begin{figure}
\centerline{\psfig{file=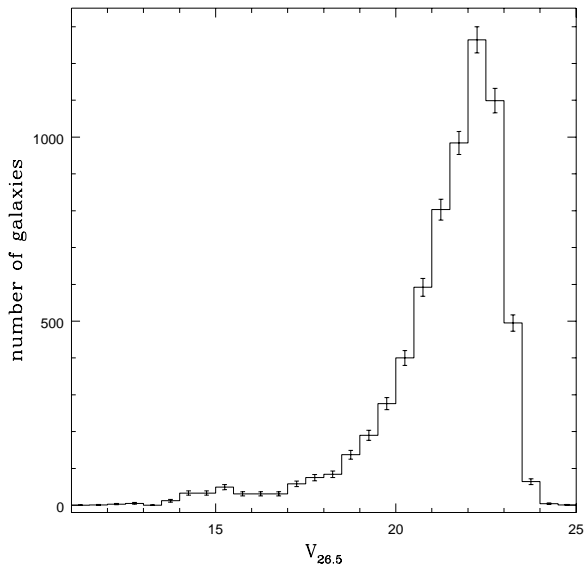,height=8.0cm}}
\caption{Histogram of magnitudes for raw galaxy counts. The turnover of 
the histogram, around V$_{26.5} \sim 22.5$, gives an estimate of the 
completeness magnitude of the observations. Poisson error bars are 
displayed.}
\protect\label{counts}
\end{figure}

Star-galaxy separation was performed based on a compactness parameter 
determined by Le F\` evre et al. (1986, see also Slezak et al. 1988). For 
each object we computed its compactness Q by~:

\begin{equation} 
Q=\frac{10^{0.4(\mu_0-V)}}{1-exp(-(r/\sigma)^2)}
\end{equation} 

\noindent where $\mu_0$, V, r and $\sigma$ are, respectively, the central 
surface 
brightness, the isophotal magnitude, the corresponding radius and the 
FWHM for that frame. By normalizing Q, we expect that its value will approach 
unity for objects with a gaussian profile, that is stars. Actually, in some 
of the cases, it will be slightly different from 1 due to a 
natural dispersion in this 
relation and to possible saturation of some of the brightest objects. The 
separation 
value (Q$_{sep}$=1.7) was then determined by eye inspection of 
the plot normalized-Q vs. V displayed in Fig.~\ref{sep}. Stars 
are expected to be placed under the y=Q$_{sep}$ line, while galaxies will be 
randomly distributed above the same line. It is evident from that same figure 
that the stellar sequence with V $<$ 15 presents Q $>$ Q$_{sep}$ but these 
are the saturated objects that were carefully flagged by visual inspection 
and classified as stars {\it a priori}.
After separation, stars represent 
approximately 35~$\%$ of the total sample, and 36~$\%$ if we restrict the 
sample to V $\leq$ 22.5, which is the completeness magnitude of our data 
as estimated by the turnover of the raw counts (see Fig~\ref{counts}).

A pitfall of this classification procedure could be the 
misclassification of compact galaxies as stars. In order to test the 
reliability of the separation, we carried out a simple test. After having 
transformed our CCD coordinates into the GMP reference system (see 
section~\ref{xy}), we identified our objects with those belonging to the 
Coma redshift catalogue obtained by Biviano et al. (1995). We thus 
estimated that, out of 278 identifications, less than 2~$\%$ of the objects 
classified as galaxies by our procedure actually had star-like velocities.

It is obvious that this test is limited to a small number of identifications, 
since we can only apply it to objects with V $>$ 15, due to saturation, and 
V~$\la$~17, which is the 95~$\%$ completeness limit of the redshift 
catalogue. Nevertheless, 
it gives a representative result for the whole sample and reassures 
us on the efficiency and accurateness of the distinguishing procedure.

After elimination of repeated detections of some of the objects (see 
section~\ref{xy}) we ended up with a catalogue containing 7023 galaxies and 
4096 stars.

\subsection{Surface brightness selection effects}

\begin{figure}
\centerline{\psfig{file=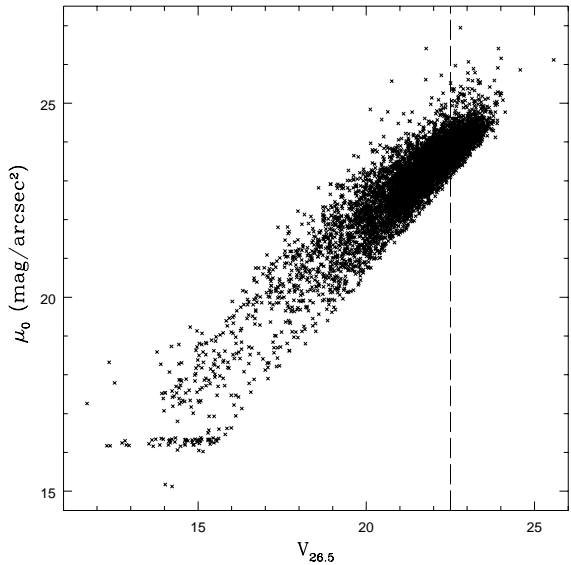,height=8.0cm}}
\caption{Central surface brightness plotted against isophotal magnitudes for 
all the catalogued galaxies. The dashed line indicates the completeness 
magnitude limit.}
\protect\label{sb}
\end{figure}

In order to test the detection limit of our observations imposed by the 
surface brightness we plot $\mu_0$ vs. V in Fig.~\ref{sb}. 
In this plot the diagonal cut shows the sequence of compact objects. 
Practically all objects below completeness magnitude 22.5 are placed at 
$\mu_0$~$\la$~24.5 mag/arcsec$^2$, as confirmed 
by the turnover value of the histogram of $\mu_0$. Above that value 
detections are sparse. 
This limiting detection value might make us miss some very faint surface 
brightness objects, but below it we estimate our catalogue to be complete 
in surface brightness.

\subsection{Zero point accuracy and errors estimated for the photometry 
derived parameters}\label{zeropoint}

\begin{figure}
\centerline{\psfig{file=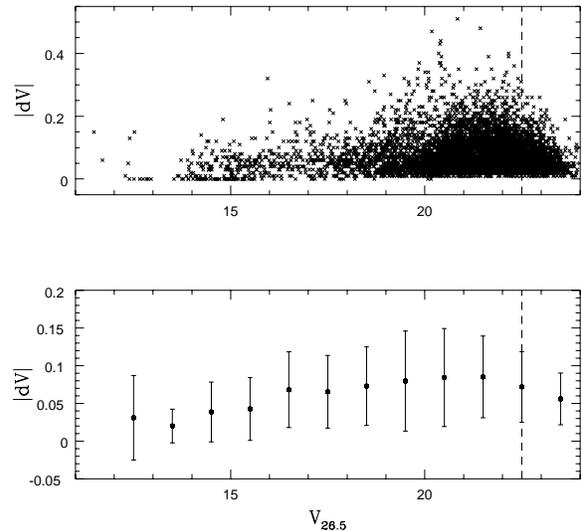,height=8.0cm}}
\caption{Magnitude total errors for all catalogued objects computed 
by means of equation~\ref{eq:dV} (upper panel). We also show, per magnitude 
bin, the mean error and dispersion (lower panel).}
\protect\label{calcmags}
\end{figure}

The estimate of magnitude errors is done frame by frame, according 
to the variations 
detected in the sky flux for each exposure. By doing so we are certain of 
estimating a total error that includes both internal errors inherent to the 
measurement algorithm, as well as external errors produced by the 
observational conditions such as differential absorption in the different 
nights of the run. We compute, for all of the objects 
in a given frame, a typical measure of the magnitude error that is given by~:

\begin{equation}\label{eq:dV} 
|dV|=-2.5~log~(flux)+2.5~log~(flux+\sigma)
\end{equation} 

\noindent where the first term of the right-hand side of the equation is the 
magnitude in the catalogue. In the second term, $\sigma$ has been computed by averaging, for each 
frame, different values of the standard deviation of the sky flux measured in 
different regions devoid of objects in that frame, and scaling the result to 
the surface of each object. The errors introduced by the flat--field procedure 
(large scale residuals) are less than 0.3$\%$, and this factor was neglected 
in the standard deviation estimation of the flux measurement.\\
In Fig.~\ref{calcmags} we plot $|dV|$ vs. V for all the objects individually 
(upper pannel) and its mean value and dispersion per magnitude bin in the 
lower pannel. The mean value is below 0.1 magnitude.

\begin{figure}[]
\centerline{\psfig{file=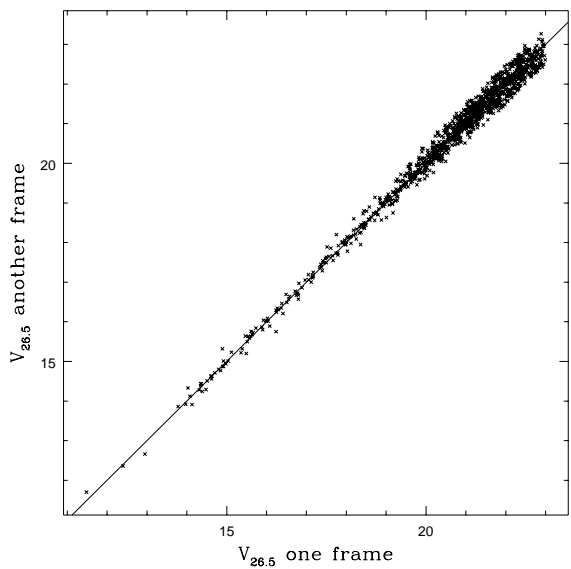,height=8.0cm}}
\caption{Galaxies measured twice~: for each one we compare its magnitude 
as obtained in two different frames.}
\protect\label{compmags}
\end{figure}

\begin{figure}[]
\centerline{\psfig{file=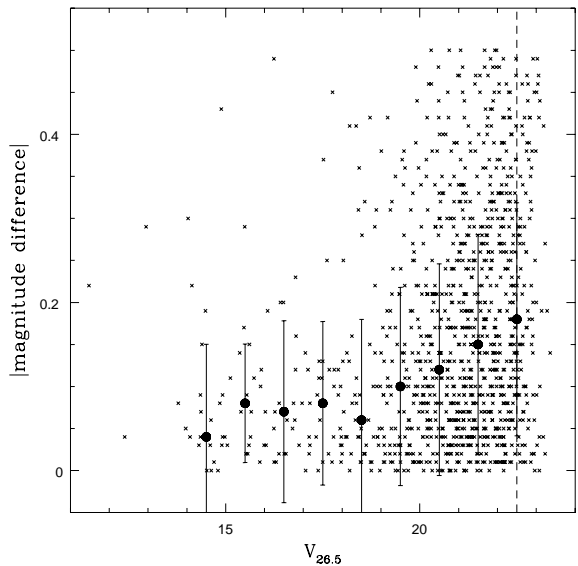,height=8.0cm}}
\caption{Crosses give the modulus of the difference 
between both magnitudes plotted in Fig.~\ref{compmags}, for each one of 
the double measured galaxies, as a function of the galaxy's magnitude as measured in one of the frames. Circles stand for the median value of this 
difference in each magnitude bin, and error bars show the dispersion around 
that median.}
\protect\label{difmags}
\end{figure}

Another point we set out to deal with in this section - zero point 
variations - is tackled by means of the 1082 galaxies with V brighter than 
the completeness magnitude value that were measured twice in the overlapping 
CCD areas (see section~\ref{xy}). Fig.~\ref{compmags} displays 
magnitudes for these objects. The points cluster 
closely around the quadrant line y=x with a larger dispersion for fainter 
magnitudes, as expected. One can notice that differences are not 
systematic. In Fig.~\ref{difmags} we quantify these 
results by computing, for each of the 1082 galaxies, the modulus of the 
difference between the magnitudes measured in two distinct frames (that is, 
the values plotted in the 2 axis of the previous figure). We also display, 
for each magnitude bin, the median and 
dispersion of those absolute differences for the objects 
belonging to that bin. Below completeness magnitude the median does not 
exceed 0.15 and one should bear in mind that this value comprises the 
magnitude errors (discussed above) for both measures. It is thus by far an 
overestimate of the zero point accuracy.

In what concerns $\mu_0$, errors range from 0.02 to 0.4 mag/arcsec$^2$ for 
bright to faint objects below the completeness magnitude limit.

\section{Astrometry}\label{xy}
We performed a standard transformation on our CCD coordinates to 
the reference system defined by Godwin et al. (1983, GMP). In this 
system each 
object has (X,Y) coordinates in arcsec, given relatively to a center 
defined at $\alpha = 12^h57.3^m, \delta=28^o14.4'~(1950)$ - located
 between both largest/brightest central galaxies, NGC~4874 and NGC~4889 
(see Fig.~\ref{map}).
For spectroscopic purposes, the frames were taken with a large 
superposition in Y (of the order of 40~$\%$, while almost negligible in
 X), which caused double observation of many objects. 
We carefully eliminated these double entries, both for stars and galaxies.
In order to estimate the final precision of our positions, we compared our 
star catalogue with the Guide Star Catalogue (GSC) of the Hubble Space 
Telescope limited to magnitude m$<$15.5, which has a 0.3 arcsec 
accuracy. The positions of the same objects in both catalogues coincided, 
after final tuning, within less than 3 arcsec (median result for 17 stars 
identified in the field of our observations).

\section{The catalogue}

The catalogue, available in electronic form alone at the CDS (Centre de 
Donn\' ees Astronomiques de Strasbourg), presents the following 
entries for all objects in both regions of observation (the large central 
zone and the smaller south-west area - see Fig.~\ref{map})~:\\
(1,2,3) Right ascension (1950).\\
(4,5,6) Declination (1950).\\
(7) Isophotal radius r$_{26.5}$ in arcsec.\\
(8) Central surface brightness $\mu_0$ in mag/arcsec$^2$.\\
(9) Isophotal apparent magnitude V$_{26.5}$.\\
(10) Magnitude error (in modulus), as estimated by equation \ref{eq:dV}.\\
(11) Classification of the object~: 1~$=$~star, 0~$=$~galaxy.\\
(12) GMP number, when available. The letter after the number indicates the GMP 
catalogue used for the matching~: g stands for the GMP (1983) galaxy 
catalogue, while s stands for the GMP unpublished star catalogue. The 
correspondance was obtained by cross-correlating positions. Do notice there 
are inevitable ambiguities in this cross-identification, possibly due to non 
resolution of close objects by GMP or simply to confusion when several close 
neighbours exist. That is why there are 4 entries doubled (because each one 
of those 4 objects was identified with 2 different GMP objects~: 3325g, 3336g, 
2976g, 2980g, 4068g, 4075g and 4032s, 4042s). In addition, there are also 
several other entries which are attributed the same GMP number. We thus 
caution the reader/potential user of this catalogue to beware not to use these 
data directly without taking into account this information.\\
(13,14) GMP (X,Y) coordinates in arcsec (after slight correction to match GSC 
positions, as described in section~\ref{xy}).\\
(15) Heliocentric radial velocity in km s$^{-1}$, when available, as given by 
Biviano et al. (1995).\\

Some of the results derived from the analysis of this catalogue are discussed 
in Lobo et al. 1996 (in press), Gerbal et al. 1996 (submitted) and the 
corresponding available spectral data is published by Biviano et al. (1995).

\acknowledgements {We thank St\' ephane Arnouts, Isabel M\' arquez, Cl\' audia 
Mendes de Oliveira and Christopher Willmer for useful discussions on 
photometry; Francois S\` evre, Jos\' e Donas and Roland den Hartog for 
astrometric discussions, and also 
J.G. Godwin, N. Metcalfe \& J.V. Peach for providing us with
their unpublished catalogue of stellar objects in the Coma field. We would 
also like to thank the referee, G. Gavazzi, for helpful comments on the text. 
We acknowledge financial support from GDR Cosmologie, CNRS; 
CL is fully supported by the BD/2772/93RM grant attributed by JNICT, Portugal.}

\end{document}